\documentclass[structabstract]{aa}  
\usepackage{graphicx}
\usepackage{natbib}
\usepackage{txfonts}
\usepackage{url}
\usepackage{color}

\newcommand{\hr}{\hbox{$^\mathrm{h}$}}
\newcommand{\m}{\hbox{$^\mathrm{m}$}}
\newcommand{\s}{\hbox{$^\mathrm{s}$}}
\newcommand{\airac}{\hbox{$\alpha_\mathrm{IRAC}$}}

\newcommand{\um}{\hbox{\,$\mu$m}}
\begin{document}
   \title{\emph{Spitzer} observations of NGC\,2264:\\The nature of the disk population}

   \subtitle{}

   \author{P.~S. Teixeira\inst{1}\thanks{ currently at the Institute for Astronomy (IfA), University of Vienna, T\"urkenschanzstrasse 17, A-1180 Vienna, Austria; paula.teixeira@univie.ac.at} 
           \and 
           C.~J. Lada\inst{2}
           \and
           M. Marengo\inst{3}
          \and
          E. A. Lada\inst{4}}

   \institute{European Southern Observatory (ESO),
              Karl-Schwarzschild-Strasse 2, D-85748 Garching b. Muenchen, Germany \\
              \and
              Harvard-Smithsonian Center for Astrophysics,
              60 Garden Street, M.S. 71, 02138 Cambridge, MA, USA\\
              \email{clada@cfa.harvard.edu}
              \and
              Department of Physics and Astronomy, Iowa State
              University, Ames, IA, USA\\
              \email{mmarengo@iastate.edu}
              \and
              Department of Astronomy, University of Florida,
              Gainesville, FL, USA\\
              \email{lada@astro.ufl.edu}
}

   \date{...}

 
  \abstract
{}
   {NGC\,2264 is a young cluster with a rich circumstellar disk population which makes it an ideal target for studying the evolution of stellar clusters. Our goal is to study the star formation history of NGC\,2264 and to analyse the primordial disk evolution of its members.}
   {The study presented is based on data obtained with the Infrared Array Camera (IRAC) and the Multiband Imaging Photometer for \emph{Spitzer} (MIPS) on board the \emph{Spitzer} Space Telescope, combined with deep near-infrared (NIR) ground-based FLAMINGOS imaging and previously published optical data.}
   {We build NIR dust extinction maps of the molecular cloud associated with the cluster, and determine it to have a mass of 2.1$\times$10$^3$\,M$_\odot$ above an A$_\mathrm{V}$ of 7\,mag. Using a differential K$_s$-band luminosity function (KLF) of the cluster, we estimate the size of the population of NGC\,2264, within the area observed by FLAMINGOS, to be 1436$\pm$242 members. The star formation efficiency is $\geq \sim$25\%. 
 We identify the disk population and divide it into 3 groups based on their spectral energy distribution slopes from 3.6\um\ to 8\um\ and on the 24\um\ excess emission: (i) optically thick inner disks, (ii) anaemic inner disks, and (iii) disks with inner holes, or transition disks.
We analyse the spatial distribution of these sources and find that sources with thick disks segregate into sub-clusterings, whereas sources with anaemic disks do not. Furthermore, sources with anaemic disks are found to be unembedded (i.e., with A$_V <$3\,mag), whereas the clustered sources with thick disks are still embedded within the parental cloud. }
   {NGC\,2264 has undergone more than one star-forming event, where the anaemic and extincted thick disk population appear to have formed in separate episodes: the sources with anaemic disks are more evolved and have had time to disperse and populate a halo of the cluster. We also find tentative evidence of triggered star-formation in the Fox Fur Nebula. In terms of disk evolution, our findings support the emerging disk evolution paradigm of two distinct evolutionary paths for primordial optically thick disks: a homologous one where the disk emission decreases uniformly at NIR and mid-infrared (MIR) wavelengths, and a radially differential one where the emission from the inner region of the disk decreases more rapidly than from the outer region (forming transition disks).}


   \maketitle
%

\section{Introduction}
\label{sec:intro}
  
The \emph{Spitzer} Space Telescope \citep{werner04}, a space-borne observatory of unprecedented sensitivity and spatial resolution in the infrared, has brought forth an exciting new era in the study of star formation. In particular, the Multiband Imaging Photometer for \emph{Spitzer} (MIPS) \citep{rieke04} has excelled in identifying protostars \citep[e.g.][]{muzerolle04,teixeira06,muench07,winston07}, and the Infrared Array Camera (IRAC) \citep{fazio04a} has permitted the detailed study of jets \citep[e.g.][]{noriega-crespo04b,teixeira08a} and circumstellar disks \citep[e.g.][]{sicilia-aguilar06,hernandez08,luhman08,sung09,sung10}. One of the earliest \emph{Spitzer} studies of circumstellar disks was that of \citet{lada06a},  who conducted a detailed study of the frequency and nature of circumstellar disks in IC\,348. The study found that this 2-3\,Myr old cluster has a disk fraction of 50\%$\pm$6\%, and that this frequency is a function of spectral type, where K6-M2 stars possess the greatest disk longevity. Regarding the nature of the disks, \citet{lada06a} identified optically thick and optically thin or anaemic  inner disks according to their spectral energy distribution (SED) slope from 3.6\um\ to 8\um\ (\airac). 
The majority of the IC\,348 sources classified as classical T Tauri stars (CTTS)  (i.e., accreting according to the equivalent width of their H$\alpha$ emission) were found to possess thick disks,  while the majority of the weak-line T Tauri stars (WTTS) (i.e., non-accreting according to the equivalent width of their H$\alpha$ emission) were found to be disk-less, leading the authors to conclude that the gaseous and dust components of disks evolve on similar timescales.

NGC\,2264 is a young hierarchical cluster \citep{piche93,lada93,lada03} located in the Monoceros OB1 association, that has been extensively studied since \citet{walker56}'s seminal work identifiying its pre-main sequence population (PMS). The most recent distance measurement to the cluster places it at 913 $\pm$110\,pc \citep{baxter09} (the value used in this paper); previous estimates of the cluster distance had varied between 700\,pc and 800\,pc.
The oldest members of the cluster are $\sim$5\,Myr old \citep{dahm05a}, although there is significant observational evidence of ongoing star formation such as CO outflows \citep[e.g.][]{margulis88}, Herbig-Haro objects \citep{reipurth04b}, and luminous far-infrared sources detected by the Infrared Astronomical Satellite \citep{margulis89}. The PMS population has furthermore been associated with  X-ray emission \citep{flaccomio06} and active accretion \citep[H$\alpha$ observations,][]{dahm05a,furesz06}. NGC\,2264 is one of the classical target clusters that has time and time again given valuable insights into the star formation process of low- and intermediate-mass stars. Initial results obtained from \emph{Spitzer} observations of NGC\,2264 led to the discovery of the Spokes cluster \citep{teixeira06}, where the nearest neighbour separations between its class\,I members are consistent with thermal Jeans fragmentation of the parental filamentary cloud. Within the Spokes cluster, a dense micro-cluster of protostars was also revealed \citep{young06}, and follow-up observations with the Submillimeter Array \citep{teixeira07} found new Class\,0 sources whose average separations were also consistent with the Jeans length for the parent molecular core of the micro-cluster. More recently, \citet{sung09} and \citet{sung10} reported the spatial distributions of Class\,I vs. Class\,II sources; they confirmed the spatial concentration of Class\,I sources in the Spokes cluster and identified another grouping of Class\,I sources near IRS\,1, that coincides with a clustering of compact millimetre sources found by \citet{peretto06,peretto07}. The northern region of NGC\,2264, near the O7 star S\,Mon, is mostly comprised of Class\,II sources; \citet{sung09} and \citet{sung10} found that the fraction of primordial disks increases with distance from S\,Mon, indicating that UV radiation plays an important role in disk evolution. The cluster has also been observed by the space-borne telescope COROT, which provided exquisitely detailed light-curves \citep{alencar10}. Comparison of the three different types of identified light curves (AA\,Tau-like, spot-like, and irregular) with \emph{Spitzer}  IRAC data showed remarkable correlation and led \citet{alencar10} to conclude that $\sim$30 to 40\% of sources with inner disks are associated with AA\,Tau-like light curves.
In this paper, we revisit the NGC\,2264 cluster by characterising the disk population and correlating its properties with those of the molecular cloud from which it formed.


\section{Data}
\label{sec:data}

\subsection{\emph{Spitzer} IRAC and MIPS data}
\label{subsec:iracdata}

For a description of the acquisition of the IRAC data used, we refer the reader to \citet{teixeira06} and \citet{young06}. 
The source extractor SExtractor software \citep{bertin96} was used to generate a list of point sources for each wavelength. Aperture photometry was performed on these detected sources using the Image Reduction and Analysis Facility (IRAF) routine APPHOT, with an aperture radius of 2 re-sampled IRAC pixels, and a sky annulus of inner and outer radii of 2 and 6 re-sampled IRAC pixels, respectively. The zero point magnitudes used for the IRAC pass-bands of effective wavelengths 3.6\um, 4.5\um, 5.8\um, and 8\um\ were 17.83, 17.34, 16.75, and 15.93 mag, respectively.
IRAC band mer\-ging was performed using the IDL function srcor \footnote{from IDL Astronomy User's Library: \url{http://idlastro.gsfc.nasa.gov/homepage.html}, adapted from software from the Ultraviolet Imaging Telescope.}, where we used a matching radius of 1 pixel (0.9\arcsec). The photometry of bright saturated sources extracted from the 10.4\,s integration images were replaced by the photometry obtained from the short-exposure, 0.4\,s integration images.
The IRAC data used in this paper was limited to photometry with errors smaller than 0.1\,mag. The completeness limits are 13.25\,mag for the 3.6\um, 4.5\um, and 5.8\um\ IRAC passbands, and 12.75\,mag for 8\um\ IRAC passband. The errors are systematically larger for the 5.8\um\ and, especially, the 8\um\ band photometry.
For a description of the MIPS data used (acquisition, calibration, and photometry), we refer the reader to \citet{teixeira06} and \citet{young06}. The completeness limit of the MIPS 24\um\ data is 7.0\,mag; we do not use the 70\um\ or 160\um\ data in this paper as very few pre-main sequence sources are detected at these longer wavelengths.
We merged these data with the IRAC sample using the same method described above for the IRAC band-merging, using the closest match within a radius of 3.5\arcsec.

\subsection{FLAMINGOS imaging}
\label{subsec:flamingos}

Deep near-infrared (NIR) $J$ (1.25\um), $H$ (1.65\um), and $K$ (2.2\um) data was obtained with the FLoridA Multi-object Imaging Near-IR Grism Observational Spectrometer \citep[FLAMINGOS\footnote{see  \url{http://flamingos.astro.ufl.edu}}; ][]{elston98, levine06} mounted on the Kitt Peak National Observatory (NOAO) 2.1\,m. The data is part of the FLAMINGOS giant molecular cloud survey\footnote{NOAO   Survey Program 2000B-0028: Towards a Complete Near-Infrared   Spectroscopic Survey of Giant Molecular Clouds (PI:  E. Lada)}. FLAMINGOS employs a 2048$\times$2048 HgCdTe HAWAII-2 infrared array with 18\um\ pixels. This corresponds to a plate scale of 0.608\arcsec\,pixel$^{-1}$ and a field of view of 20.5\arcmin$\times$20.5\arcmin. Table \ref{tab:flamingos} summarises the observing information for the fields acquired in NGC\,2264. The FLAMINGOS observations cover an area of $\approx$ 0.3 square degrees. The seeing varies within the dataset, at all wavelengths, from 0.9 to $\approx$1.5\arcsec\ FWHM. The data reduction, calibration, and photometry was performed with custom FLAMINGOS data analysis and reduction pipelines within IRAF. For a detailed description of these, we refer the reader to \citet{roman-zuniga08}. 

We compared the FLAMINGOS and Two Micron All Sky Survey  (2MASS) \citep{kleinmann94,skrutskie06} data and identified the sources with saturated FLAMINGOS photometry (the bright sources have FLAMINGOS photometry that is systematically fainter then their corresponding 2MASS photometry). These saturated sources had their FLAMINGOS photometry replaced by 2MASS photometry. The saturation limits for FLAMINGOS derived from this comparison are 11.0, 11.5, and 12.0\,mag for $J$, $H$, and $K$, respectively.

The area covered in NGC\,2264 by the FLAMINGOS survey is smaller than that covered by the IRAC observations. We therefore made use of 2MASS photometry to complete the NIR coverage of the region observed by IRAC. The NIR data are thus not uniform in depth since the FLAMINGOS data has completeness limits of 19.0, 18.5, and 17.5\,mag for $J$, $H$, and $K_s$, respectively, while the shallower 2MASS data have corresponding completeness limits of 15.5, 14.5, and 13.5\,mag. The analysis presented in this paper is based on NIR photometry that is within the completeness limits and has photometric errors smaller than 0.1\,mag.

\subsection{Ancillary data}
\label{subsec:ancillary}

To perform a more complete study of NGC\,2264, we used previously published datasets, in particular the ancillary data catalogue (ADC) built by \citet{rebull06}, which is an extensive collection of multiwavelength photometric and spectroscopic data of NGC\,2264. 

We added additional photometric and spectroscopic datasets to the ADC. One of the additional datasets comprises the data on NGC\,2264 collected by \citet{meyers-rice95}; these consist of $VRI$ photometry and optical spectroscopy. Spectra of 361 stars were obtained in January 1993 using the Boller \& Chivens spectrograph (3700\,\AA\ to 7000\,\AA) with the Steward Observatory 2.3\,m Bok telescope at Kitt Peak, Arizona. A first set of imaging data were acquired during November 1989 and January 1990 using the Bok telescope and a 800\,$\times$\,800 pixel CCD, and a second set was obtained in December 1992 with the same telescope but using the Loral/Fairchild camera (2048\,$\times$\,2048; the data obtained in the second set was binned to match the plate scale of the first set). The photometric data were calibrated using data for the globular clusters NGC\,7790 and NGC\,2419.

An X-ray study of NGC\,2264 was conducted by \citet{flaccomio06} u\-sing ACIS-I  onboard the Chandra X-ray Observatory. The data obtained has a total integration time of 97\,ks (observation ID 2540; general observer proposal PI S. Sciortino), and the area observed covers 17\arcmin\,$\times$\,17\arcmin\ centred on \hbox{$(\alpha,\delta)(J2000)$=(06\hr40\m58.7\s, +09\degr34\arcmin14\arcsec)} resulting in the detection of 420 sources. These data were subsequently added to the ADC. 

The final dataset added corresponds to that published by \citet{furesz06}. These authors performed a radial velocity survey of NGC\,2264 by obtaining spectroscopic observations of 990 optically visible stars selected from the studies described above. The data were obtained using the Hectochelle, mounted on the 6.5\,m MMT in Mount Hopkins, Arizona (U.S.A). \citet{furesz06} classified 471 stars as cluster members (according to criteria based on their radial velocities and/or H$\alpha$ line profiles), of which 344 have well-determined radial velocities. 

The analysis we present in this paper was done using a subset of this final merged catalogue: only sources that were detected in all four IRAC bands were considered, and we refer to this subset henceforth as the \emph{IRAC sample}.


\section{Results}
\label{sec:results}


\subsection{Identification of the disk population}
\label{subsec:disks}

To begin characterising the disk population of NGC\,2264, we uses a tool developed by \citet{lada06a} in their study of the young cluster IC\,348; they measured the slope of the SED, for each source, between 3.6\um\ and 8\um. We used this same measure, \airac, to classify the evolutionary status of sources with respect to their inner circumstellar material (emission longwards of 8\um\ is considered to arise from outer circumstellar material). The sources in the IRAC sample were classified according to the \airac\ empirical scheme shown in Table \ref{tab:class}.
Class\,I sources viewed pole-on may have flat spectrum SEDs \citep{calvet94,robitaille06}, and so may sources with optically thick circumstellar disks (CTTS or Class\,II sources) if viewed edge-on \citep{robitaille06}. We thus added an intermediate evolutionary phase between Class\,I sources and sources with thick disks to the empirical classification scheme of \citet{lada06a}: the colloquially termed \emph{flat spectrum} phase. Several authors have suggested that flat spectrum sources may indeed be a distinct evolutionary group altogether, such as \citet{greene02} who found that the accretion rates of flat spectrum sources are consistently higher than those for CTTS and lower than those for Class\,I sources, and \citet{muench07}, who found that flat spectrum sources are intrinsically more luminous than Class\,II sources.

\begin{figure}[!h]
\centering
\includegraphics[width=0.48\textwidth]{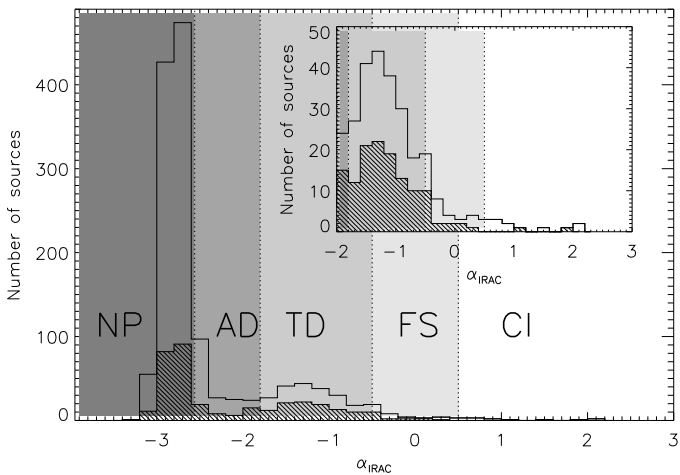}
\caption{Distribution of \airac\ for sources in the IRAC sample (open histogram) and for 349 candidate members previously identified by \citet{furesz06} (hatched-line histogram). The small inset is the same diagram but zoomed into the \airac\ region corresponding to sources with thick disks. The shaded areas correspond to the regions in the histograms occupied by Class\,I (C\,I) sources, flat spectrum (FS) sources, sources with thick disks (TD), sources with anaemic disks (AD), and sources with naked photospheres (NP).}
\label{fig:airac-hist}
\end{figure}

Figure \ref{fig:airac-hist} shows the distribution of \airac\ for 1437 IRAC sample sources (open histogram) and for 349 candidate members previously identified by \citet{furesz06} from their radial velocity survey (filled hatched-line histogram). The IRAC sample \airac\ histogram shows that the majority (985) of the sources have naked photospheres - these are comprised of disk-less PMS sources of NGC\,2264, as well as foreground and background field stars. The IRAC sample has 408 sources that have circumstellar material (\airac $>$ -2.56), 160 of which had been previously identified by \citet{furesz06} as candidate members.
Of the 349 sources, $\approx$46\% have circumstellar material, either from envelopes (there are 11 Class\,I and flat spectrum sources in \citet{furesz06}'s list that are detected in all four IRAC bands), thick (30\%), or anaemic disks (12\%). We note that the statistics for flat spectrum and Class\,I sources are particularly incomplete since these are very red sources and some may not have been detected in all four IRAC bands; the above fraction of sources with circumstellar material is thus an underestimate. 

\begin{figure}[!h]
\centering
\includegraphics[width=0.48\textwidth]{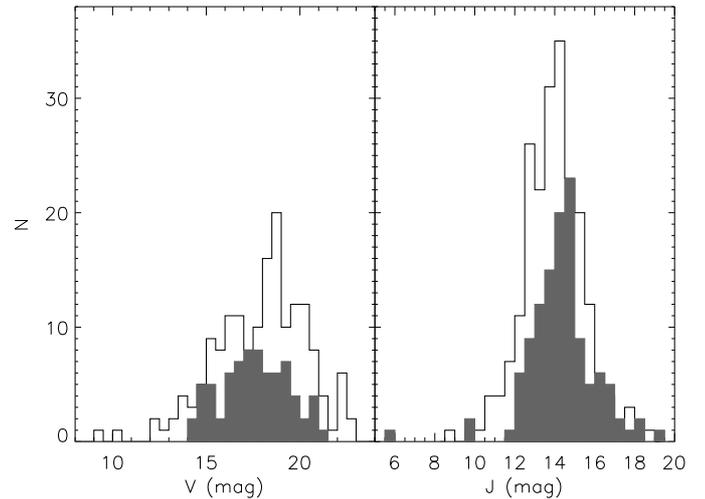}
\caption{Comparison of V- and J-band luminosity functions for sources with thick disks (open histogram) and sources with anaemic disks (filled histogram).}
\label{fig:histograms}
\end{figure}

Extragalactic sources have infrared excess emission that can be erroneously interpreted as a signature of a star with a circumstellar disk \citep[e.g.][]{jorgensen06}.
Although the extensive molecular cloud associated with the NGC\,2264 cluster \citep[e.g.][]{ridge03} is relatively efficient in screening against the contamination of extragalactic sources in the IRAC sample, it is still necessary to prune the dataset to reduce the number of these interlopers.
\citet{fazio04b} found that the contamination by extragalactic sources, such as PAH-rich galaxies, is greater than 50\% for sources fainter than 14.5 mag at 3.6\um. 
To reduce the contamination of extragalactic sources, we removed 8 sources from our IRAC sample that are fainter than 14.5 mag at 3.6\um. This extragalactic decontamination tech\-nique has also been used by other authors to filter their data \citep[e.g.][in their study of the young $\sigma$ Orionis cluster]{hernandez07}. Figure \ref{fig:histograms} shows that the luminosity functions for V- and J-bands are similar for sources with thick disks (open histogram), and sources with anaemic disks (filled histogram). Since the thick and anaemic disk populations have consistent luminosity functions, we conclude that the contamination of extragalactic sources in the anaemic disk population is very low. Futhermore,  we found that the average radial velocity of sources with anaemic disks is 18.9$\pm$2.8\,km\,s$^{-1}$, while sources with thick disks have an average radial velocity of 20.0$\pm$4.6\,km\,s$^{-1}$. The consistency of the average radial velocities of the disk populations also led us to conclude that there is little contamination by foreground and background stars. 

The cleaned IRAC sample, used henceforth, thus has 1401 sources of which 12 are Class\,I sources, 28 are flat spectrum sources, 214 are sources with thick disks, and 154 are classified as sources with anaemic disks. Disk-less sources that are heavily reddened have \airac\ values that would lead to their identification as sources with anaemic disks, as shown by \citet{muench07}. The authors used the photosphere of a K0 diskless star and created reddened SEDs for various amounts of extinction; they found that for A$_V >$10\,mag, the SED slope \airac\ of a background diskless star was greater than -2.56, and thus the reddened star could be erroneously classified as having an anaemic inner disk.
Indeed 36 such sources were found in NGC\,2264, so the number of sources with anaemic disks is actually considered to be 118.

\begin{figure}[!h]
\centering
\includegraphics[width=0.48\textwidth]{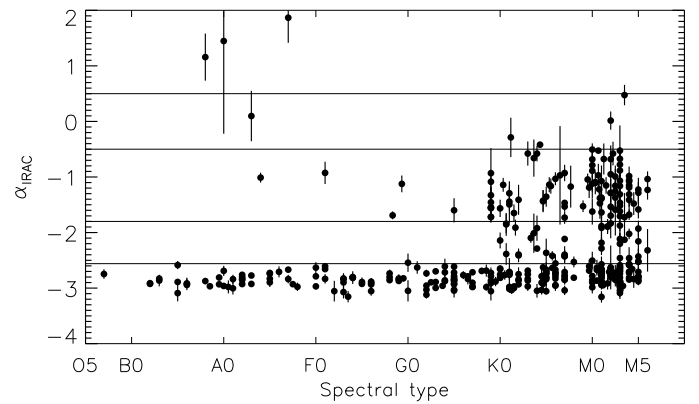}
\caption{Variation in \airac\ with spectral type. The horizontal thick lines correspond to the limits used to classify the evolutionary phases of protostars and PMS objects.}
\label{fig:spt-airac}
\end{figure}

The IRAC sample has 386 sources with previously determined spectral types. Figure \ref{fig:spt-airac} shows how \airac\ varies as a function of spectral type. As the diagram shows, the majority of sources with thick or anaemic disks are late-type ($>$G9), although some early-type sources have thick disks. A com\-pa\-ri\-son of Figure \ref{fig:spt-airac} with Figure 1 of \citet{lada06a} shows great similarity in the variation of \airac\ with spectral type between the NGC\,2264 and IC\,348 clusters. If these sources are coeval, this diagram indicates that the disks around lower mass stars may be longer lived, as suggested by \citet{lada06a}.
\subsection{The molecular cloud}
\label{subsec:cloud}

\begin{figure}[!h]
\centering
\includegraphics[height=0.48\textwidth,angle=90]{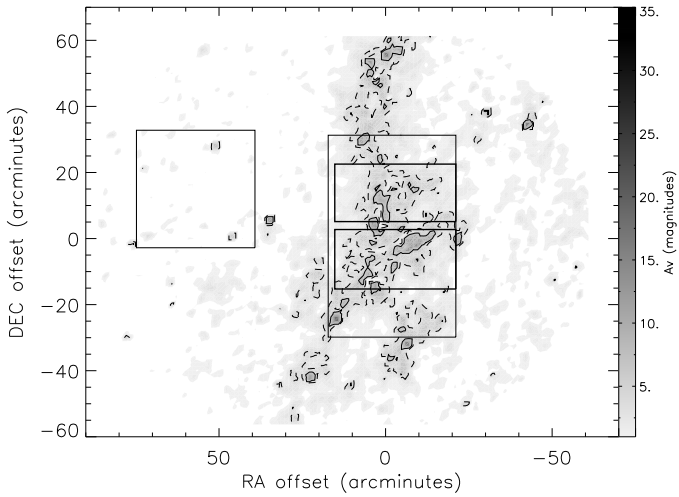}
\caption{Nyquist sampled, 2.5\arcmin\ resolution 2MASS dust extinction map of the molecular cloud associated with NGC\,2264. The map is centred on ($\alpha$, $\delta$)(J2000.0)=(6\hr41\m02.6\s, +09\degr36\arcmin12.2\arcsec). The dashed contour corresponds to A$_\mathrm{V}$=4\,mag, while the solid contour corresponds to A$_\mathrm{V}$=8\,mag. The box on the eastern part of the map indicates the size and location of the control field, while the larger box centred on the molecular cloud represents the area surveyed by IRAC. The FLAMINGOS fields, also centred on the cloud, are represented by boxes drawn with thicker lines.}
\label{fig:av-2mass}
\end{figure}

To better characterize the nature of the disk population of NGC\,2264, it is useful to  study the structure of the molecular cloud associated with the young cluster.  We began this study by building a NIR dust extinction map of the region surrounding the NGC2264 cluster, shown in Figure \ref{fig:av-2mass}. The map was built using the near-infrared colour excess (NICE) method  \citep[see][for a detailed description of the method]{lada94,alves98} and 2MASS $H$ and $K_s$ data \citep{skrutskie06}.  Basically, the NICE method measures the reddening of the ($H$-$K_s$) colour of a star background to a molecular cloud. Young stars with circumstellar disks may also have reddened ($H$-$K_s$) colours due to the reprocessed stellar emission from the dust in their inner disks - it is therefore ne\-ces\-sary to remove these disked sources from the NIR photometric sample used to build a NICE extinction map. We used the T Tauri locus in a ($H$-$K_s$) vs. ($J$-$H$) colour-colour diagram, defined by \citet{meyer97} and the \citet{cardelli89a} extinction law to remove all sources with ($H$-$K_s$) excess emission arising from the inner disk.

The visual extinction A$_\mathrm{V}$ was then calculated for each star using the following equation

\begin{equation}
A_\mathrm{V}=15.93[(H-K_s)_{observed} - (H-K_s)_{intrinsic}],
\label{eq:k}
\end{equation}

\noindent where ($H$-$K_s$)$_{observed}$ is the colour of a (disk-less) source observed towards the NGC\,2264 field and ($H$-$K_s$)$_{intrinsic}$ is the intrinsic stellar colour measured from a control field. Figure \ref{fig:av-2mass} shows the location of the control field used, which is represented by a box in the eastern part of the map. The area of the control field is the same as that of the total area covered by the FLAMINGOS observations: 1265.48 arcmin$^2$. To determine ($H$-$K_s$)$_{intrinsic}$, we used the median ($H$-$K_s$) colour of stars in the control field, i.e. 0.17\,mag. The A$_\mathrm{V}$ of each star in a pixel (2.5\arcmin\ $\times$ 2.5\arcmin\ box) was then averaged to determine the extinction for that pixel. The 2MASS dust extinction map has an average of 22 stars per pixel in the region of the molecular cloud.

Since the 2MASS data is relatively shallow, it is unable to probe the densest parts of the cloud, which is why a pixel size of  2.5\arcmin\ $\times$ 2.5\arcmin\  was used: a finer resolution (smaller pixel size) yields a map with many ``holes'' corresponding to dense regions of the cloud where 2MASS did not detect background sources to the molecular cloud (i.e., empty pixels). As a result of the lower resolution of the dust map, the average A$_\mathrm{V}$ calculated for each pixel is an underestimate to the actual extinction because the higher A$_\mathrm{V}$ values are being diluted by those given by the neighboring less-extincted stars.

\begin{figure}[!h]
\centering
\includegraphics[width=0.48\textwidth]{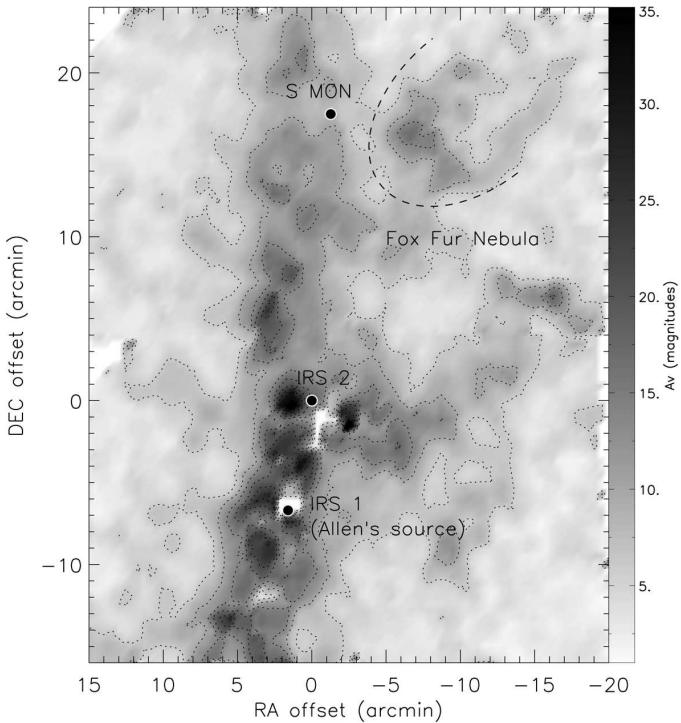}
\caption{FLAMINGOS dust extinction map of the molecular cloud associated with the NGC\,2264 cluster. The map is centred on ($\alpha$, $\delta$)(J2000.0)=(6\hr41\m02.6\s, +09\degr36\arcmin12.2\arcsec). The spatial resolution is 50\arcsec and the contours range from 6 to 30\,mag in steps of 4\,mag. For positional reference, the sources IRS\,1, IRS\,2, and S\,Mon are represented by filled circles, while the approximate southeastern edge of the Fox Fur Nebula is marked by a dashed curve.}
\label{fig:av-flamingos}
\end{figure}

Although the 2MASS extinction map is useful for an overview of the extent of the molecular cloud and surrounding region, the resolution can be improved by a factor of three with the deeper NIR FLAMINGOS data. We  therefore built a higher resolution (50\arcsec) extinction map of the molecular cloud, shown in Figure \ref{fig:av-flamingos}. As for the 2MASS dust map, we removed disk sources from the photometric catalogue prior to building the FLAMINGOS dust map. As shown in section \S\,\ref{subsec:disks}, the region covered by the FLAMINGOS data was also observed by \emph{Spitzer} IRAC; we were therefore able to use the \airac\ disk identification criterium (i.e., sources with \airac\ $>$ -2.56.) to remove these sources from the catalogue.

To reduce contamination from foreground sources (that would erroneously lower the measured value of A$_\mathrm{V}$ for each pixel in the dust extinction map), we used only sources fainter than H=15\,mag. The dust map of  Figure \ref{fig:av-flamingos} shows some ``holes'', in particular, one corresponding to the position of IRS\,1 ($\sim$ 110 southeast of the centre of the map); this hole is generated because IRS\,1 is a bright source that saturated at $K_s$ and no good photometry could be extracted for the disk-less sources in its immediate vicinity.

The densest regions in the FLAMINGOS extinction map correspond to the Spokes cluster (on which Figure \ref{fig:av-flamingos} is centred) and southwards towards IRS\,1. The mean and median visual extinction values derived from the FLAMINGOS NIR dust map are 5.9 and 4.3\,mag, respectively.
Comparison of the FLAMINGOS NIR dust map with the velocity integrated $^{13}$CO (1-0) and C$^{18}$O (1-0) maps obtained by \citet{ridge03} shows remarkable agreement with the structure of the molecular cloud. In particular, the molecular cloud component associated with the ``Fox Fur Nebula'' (a tilted ``U''-shaped feature located in the northwestern part of the cloud) is clearly recognisable in the dust extinction map. The $^{13}$CO maps of Figure \ref{fig:radio} shows that the cloud is essentially divided into two velocity components: (1) a 4-6\,km\,s$^{-1}$ component consisting of the bulk of cloud (including the Spokes cluster and the region surrounding IRS\,1), and (2) a 10-11\,km\,s$^{-1}$ component corresponding to the Fox Fur Nebula.

The mass of the molecular cloud was determined from the FLAMINGOS dust extinction map, using the equation \citep{dickman78}
\begin{equation}
M=(\alpha d)^2 \frac{N(HI+H_2)}{A_V} \mu \sum_i A_V(i),
\label{eq:av}
\end{equation}

\noindent where $\alpha$ corresponds to the pixel size of the map in radians, $d$ is the distance to the cloud, $N(HI+H_2)/A_V$=1.9\,$\times$10$^{21}$\,cm$^{-2}$\,mag$^{-1}$, the mean molecular weight, $\mu$, is 1.37, and the sum is over an individual pixel, $i$. 
The total mass obtained is M$_{total}$=3.7$^{+0.9}_{-0.8}\times$10$^3$\,M$_\odot$ (the dominant contributor to the error in mass is the uncertainty in the distance). The mass contained in regions where A$_\mathrm{V} >$\,7\,mag is M$_{A_V > 7mag}$=(2.1$\pm$0.5)$\times$10$^3$\,M$_\odot$, and the Fox Fur Nebula has M$_{Fox Fur}$= (1.2$\pm$0.3)$\times$10$^2$\,M$_\odot$.

The total mass derived from the dust extinction map is consistent with the mass determined by \citet{ridge03} from their $^{13}$CO (1-0) map, 5.2$\times$10$^3$\,M$_\odot$ (for a distance of 913\,pc), taking in account that their mass estimate is accurate to within a factor of 2- 4. Although the area covered by the FLAMINGOS dust map is 70\,\% that of the CO maps ( the $^{13}$CO map includes emission extending northwards and southwards of the dust map), our FLAMINGOS dust map covers all of the area with C$^{18}$O emission. Comparison of the mass in the dense regions of the cloud (M$_{A_V > 7mag}$) with that determined from the C$^{18}$O (1-0) map, 2.1$\times$10$^3$M$_\odot$ \citep[][for a distance of 913\,pc]{ridge03}, does indeed show good agreement.

\subsection{Extinction and spatial distribution of the disk population}
\label{subsubsec:disk-av}

Figure \ref{fig:reddening-gap} shows two NIR+IRAC colour-colour diagrams of sources in the IRAC sample. The different source classes are clearly discriminated in these colour-colour diagrams: the disk-less sources (represented by small grey dots) occupy the region corresponding to the main sequence and its reddening band. The unembedded sources with disks are located along a band of relatively constant ($J$-$H$) colour. Embedded sources with thick disks, flat spectrum sources, or Class\,I sources fan out in the NIR+IRAC diagrams at the red ($K_s$-[4.5]) or ($K_s$-[8]) end of the disk locus. These colour-colour diagrams may be used to identify and classify sources with disks (and circumstellar envelopes).  We note that the diagrams show a \emph{gap amongst the reddened sources}, an effect that is particularly noticeable in panel (b) of Figure \ref{fig:reddening-gap} - there are very few reddened sources with anaemic disks.

To quantitatively determine the disk loci, we built four NIR+IRAC colour-colour density plots that are shown in Figure \ref{fig:hess}: (a) ($K_s$-[3.6]), (b) ($K_s$-[4.5]), (c) ($K_s$-[5.8]), and (d) ($K_s$-[8]) vs. ($J$-$H$). We selected sources occupying  a region in the ($K_s$-[8]) vs. ($J$-$H$) colour-colour space with a high density ($>$10\,mag$^{-2}$) of excess emission sources to fit an empirical loci for T Tauri stars. The locus in a colour-colour diagram can be given by

\begin{equation}
y=a_0+a_1x,
\label{eq:loci}
\end{equation}

\noindent where $y$ corresponds to the ($J$-$H$) colour and $x$ corresponds to one of the four colours: ($K_s$-[3.6]), ($K_s$-[4.5]), ($K_s$-[5.8]), or ($K_s$-[8]). Table \ref{tab:loci} summarises the coefficients $a_0$ and $a_1$ obtained from the four fits. In the absence of spectral information, these loci may be used to identify embedded or unembedded sources with circumstellar disks. We compared the empirically determined locus with the colour-colour distribution of unreddened models of disked sources in the Appendix A. Using a 5-$\sigma$ limit for the disk locus of the ($K_s$-[8]) vs. ($J$-$H$) diagram, we are able to clearly separate sources with A$_V$ greater than 3\,mag. We note that previous optical studies only identified PMS sources in the cluster for A$_V <$3\,mag \citep[e.g., see Figure 8 of][]{dahm05a}, it is necessary to use longer wavelength data such as that presented in this paper \citep[and also shown in][]{sung09,sung10} to identify the embedded population.

\begin{figure}[!h]
\centering
\includegraphics[width=0.48\textwidth]{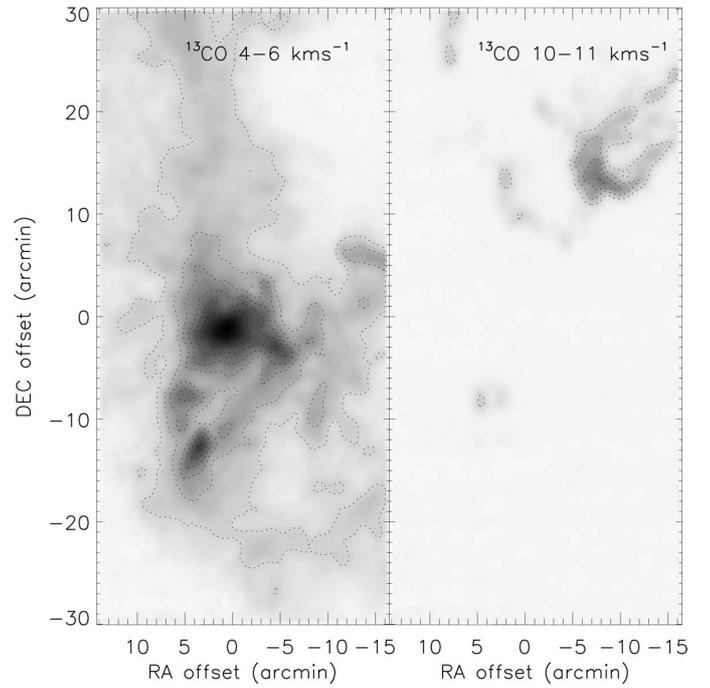}
\caption{ $^{13}$CO intensity maps, integrated for two velocity intervals, of the molecular cloud associated with NGC\,2264 cluster \citep[see][]{ridge03}. The bulk of the molecular cloud has V$_\mathrm{LSR}$ between 4\,kms$^{−1}$ and 6\,km\,s$^{−1}$, and a clear separate component is present at 10-11\,km\,s$^{−1}$ corresponding to the Fox Fur Nebula (compare with Fig. \ref{fig:av-flamingos}). The contours are the same for both panels, starting at 1.5\,K\,km\,s$^{-1}$ with increments of 1.5\,K\,km\,s$^{-1}$.}
\label{fig:radio}
\end{figure}

\begin{figure}[!h]
\centering
\includegraphics[width=0.48\textwidth]{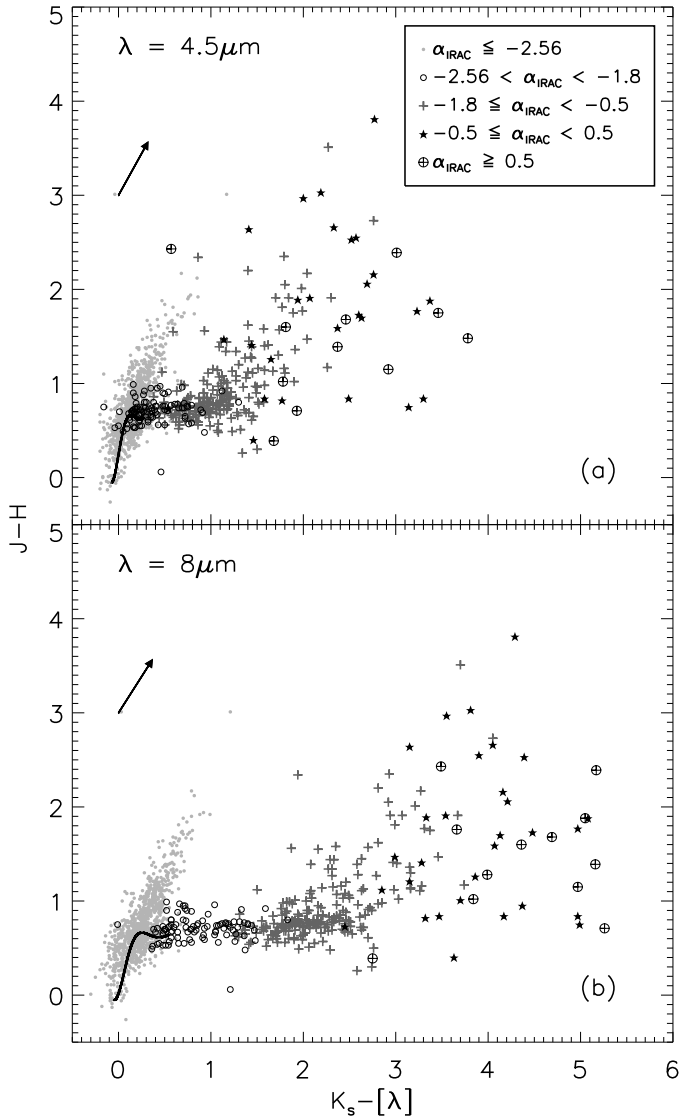}
\caption{NIR+IRAC colour-colour diagrams of sources in the IRAC sample. The symbols are as follows: disk-less sources are represented by grey dots, sources with anaemic disks by black open circles, sources with thick disks by dark grey crosses, sources with flat spectrum SEDs by filled black stars, and finally, Class\,I sources by open circles and a cross. The main sequence is plotted as a black line. The diagrams show that sources with circumstellar material can be clearly distinguished from the main sequence population. Sources with anaemic disks populate the disk loci and are not reddened, generating a reddening gap that is particularly striking in panel (b). Sources with thick disks also populate the disk loci, but some are also reddened and fan out in the diagram along with flat-spectrum sources and Class\,I sources. }
\label{fig:reddening-gap}
\end{figure}

\begin{figure}[!h]
\centering
\includegraphics[width=0.48\textwidth]{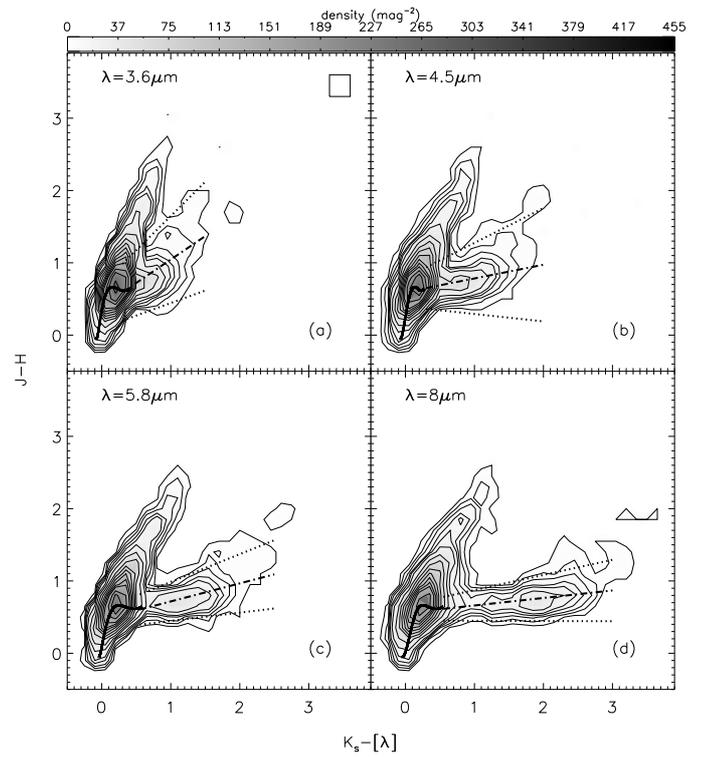}
\caption{NIR+IRAC colour-colour Nyquist-sampled density diagrams for sources in the IRAC sample. The beam size is 0.3$\times$0.3\,mag and is represented by a pixel in the upper right corner of panel (a). The contours are the same for each diagram and correspond to 3, 6.5, 10, 15, 20, 30, 50, 75, 100, 125, 150, 175, 200, 250, 300, and 400 sources mag$^{-2}$. The solid black curve corresponds to the main sequence of the Pleiades \citep[see][]{lada06a}. The dashed-dotted lines correspond to the loci of disked sources for each colour-colour space (see Table \ref{tab:loci}). The dotted lines correspond to 3-$\sigma$ (5-$\sigma$) error in the disk loci for the upper (lower) two panels.}
\label{fig:hess}
\end{figure}

\begin{figure}[!h]
\centering
\includegraphics[width=\textwidth]{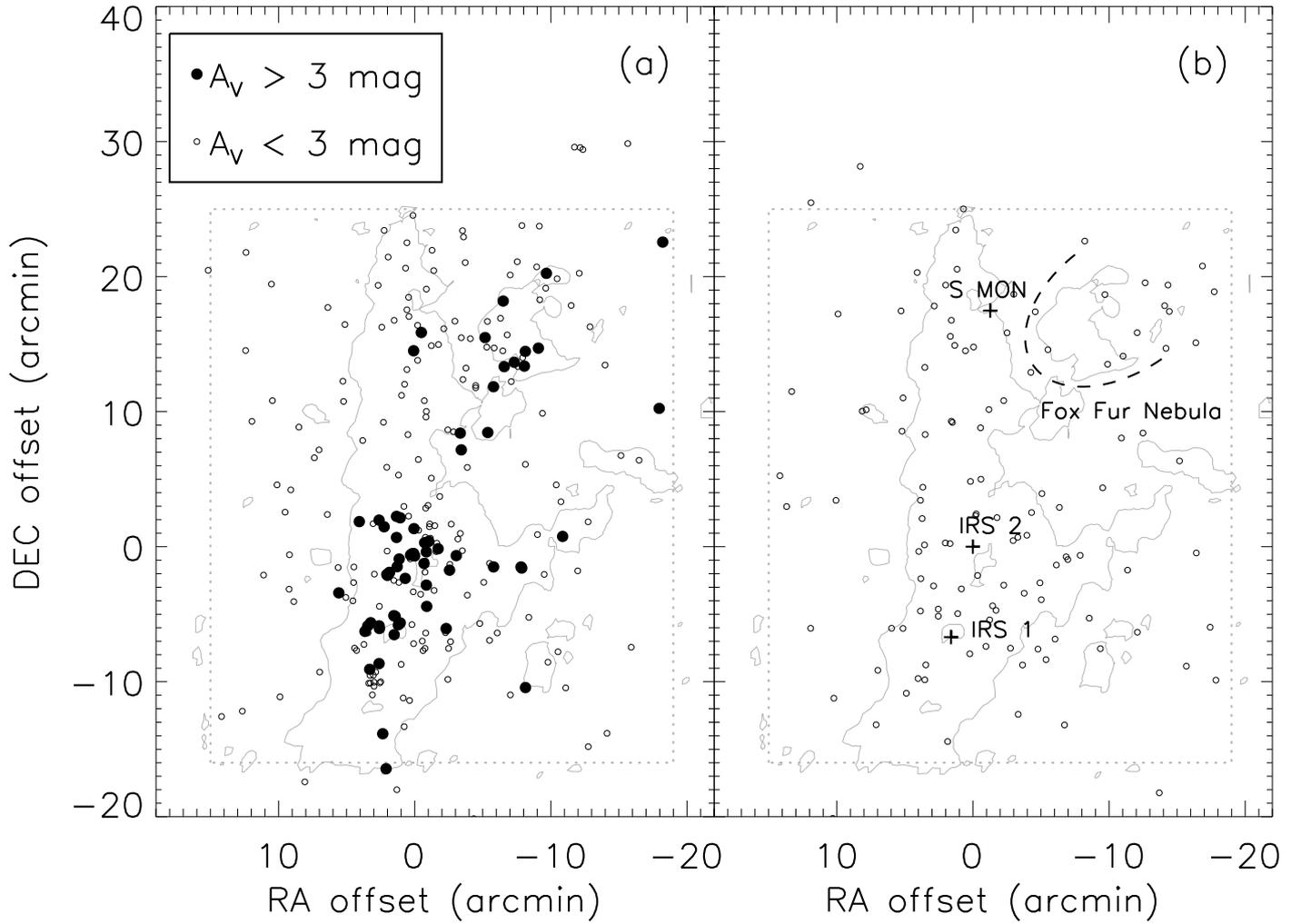}
\caption{Spatial distribution of sources with thick disks (a) and anaemic disks (b). Embedded sources (A$_V >$\,3\,mag) are represented by filled circles, and unembedded sources (A$_V \le$\,3\,mag)  are represented by open circles.  The spatial extent of the molecular cloud is represented by a grey contour of A$_\mathrm{V}$=8\,mag. Both maps are centred on IRS\,2. For positional reference, the sources IRS\,1, IRS\,2, and S\,Mon are represented by plus signs, while the approximate southeastern edge of the Fox Fur Nebula is marked by a dashed curve in panel (b).}
\label{fig:spatial}
\end{figure}

Figure \ref{fig:spatial} shows the spatial distribution of sources with thick (a) and anaemic (b) disks relative to the dense material of the molecular cloud (represented by a A$_\mathrm{V}$= 8\,mag grey contour). The thick and anaemic disk populations follow distinct spatial distributions, and also have distinct reddening distributions; the distribution of the embedded thick disk population is clearly segregated and confined to the densest regions of the cloud. \citet{sung09}'s analysis of the spatial distribution of sources with disks in NGC\,2264 lead to the identification of three clusterings: (i) near IRS\,2 or the Spokes cluster, (ii) near IRS\,1 or Allen's source, and (iii) near S\,Mon. For the sources with thick disks, groups (i) and (ii) are clearly identified in Figure \ref{fig:spatial}. We additionally note that there is a loose clustering of embedded sources with thick disks in the Fox Fur Nebula region. The clustering is small compared to those of IRS\,1 (Allen's source) and IRS\,2 (Spoke's cluster) and is likely unbound; it is composed of 19 sources with thick disks, 9 of which are embedded.  \citet{ridge03}'s $^{13}$CO observations of the molecular cloud show that the bulk of the cloud has V$_\mathrm{LSR}$=4-6\,km\,s$^{-1}$ and that there is a separate component of  V$_\mathrm{LSR}$=10-11\,km\,s$^{-1}$ corresponding to the Fox Fur Nebula (see Figure \ref{fig:radio}). As this component collides with the bulk of the cloud, it could increase the local densities and subsequently  trigger star formation, which could explain the clustering of embedded sources with the thick disks we identify here. 

Panel (b) of  Figure \ref{fig:spatial}  shows that most of these sources lie within the contour of the dense cloud material, A$_V$=8\,mag, but otherwise there is no spatial segregation between sources with anaemic disks. Furthermore,  sources with anaemic disks are not embedded, while sources with thick disks that are clustered (i.e.,  in the Spokes cluster, near Allen's source, and in the vicinity of the Fox Fur Nebula) are all embedded. \citet{sung09} also defined a more dispersed halo population, more uniformly spread out in the cluster - its spatial distribution is consistent with that of the anaemic disk population.

\subsection{Accretion activity of the disk population}
\label{subsubsec:airac-halpha}

\begin{figure}[!h]
\centering
\includegraphics[width=0.48\textwidth]{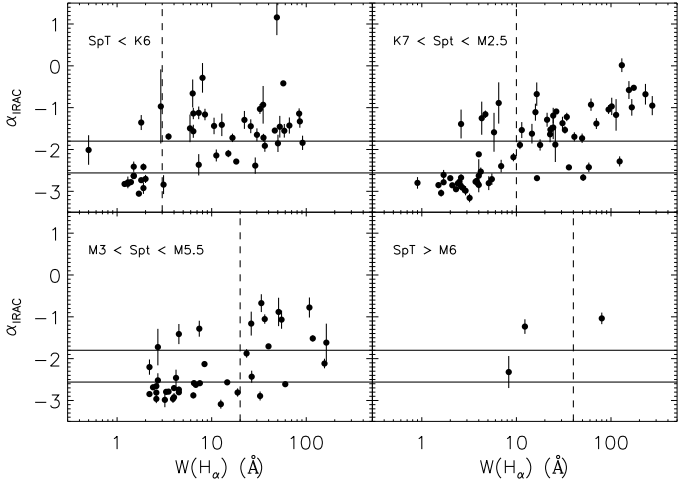}
\caption{Comparison of the H$\alpha$ equivalent width (WH$\alpha$)  with \airac\ for sources of different spectral type (SpT). The horizontal lines mark the separation between disk-less sources, sources with anaemic, and sources with thick disks (see Table \ref{tab:class}), while the vertical dashed lines separate accreting and non-accreting sources.}
\label{fig:airac-halpha}
\end{figure}

To investigate disk accretion, we consider a subsample of H$\alpha$ emitting sources \citep{dahm05a} from our IRAC sample; this subsample consists of 172 sources with measured H$\alpha$ equivalent widths, W(H$\alpha$).  Figure \ref{fig:airac-halpha} shows how W(H$\alpha$) varies for stars of different \airac\ values; the figure is composed of four panels corresponding to stars of four spectral type groups. The W(H$\alpha$) limit used to classify a source as accreting or non-accreting is a function of spectral type \citep{white03}. These different limits are represented by vertical dashed lines in Figure \ref{fig:airac-halpha}; sources located to the right of these lines are considered to be accreting.

The diagrams show that the majority ($\sim$85\%) of the sources with thick disks appear to be accreting (consistent with what \citet{lada06a} found in the young cluster IC\,348), in comparison, 59\% of sources with anaemic disks are accreting. Figure \ref{fig:airac-halpha} also shows the existence of sources with passive anaemic disks and sources  with no IRAC disk that are non-accreting - these are expected to occur if the gas and dust removal in the disk occur at approximately the same rate.  A particularly interesting region in the diagrams corresponds to that inhabited by accreting  sources with no IRAC excess emission. 
There are 17 sources with anaemic IRAC disks and 4 sources with no IRAC or inner disk that are accreting. Of the accreting sources with anaemic IRAC disks, six have excess 24\um\ detections indicating the existence of thick outer disks, and fifteen of them have X-ray emission. Among the accreting sources  with no IRAC disk, all four of them are X-ray emitting sources and three of them possess excess 24\um\ emission characteristic of a thick outer disk; the present data does not exclude the possibility that the other sources may also possess a thick outer disk. The X-ray sources with no IRAC disk and excess 24\um\ emission must therefore each have an inner hole that does not contain dust, but does contain some gas as it continues to accrete; the accretion is indicated by the large value of W(H$\alpha$) and excess UV.  We classify these types of sources as having transition disks, i.e., \emph{disks with inner dust holes filled with gas, and optically thick outer dust+gas disks}. Accreting sources with anaemic inner disks and thick outer disks could be the immediate precursors to these transition disks. 
 At this point, we remind the reader that the nomenclature for disk classification is still evolving and as such, non-standardized; a summary of the different terminologies used by researchers can be found in \citet{evans09}. In particular, some authors would consider the disks that we define in this paper as anaemic to also be transition disks, because they are more evolved \citep[e.g.][]{najita07b,currie09,currie11}. While some of the sources with anaemic IRAC disks that we identify in NGC\,2264 may be precursors to sources with disks that have cleared-out inner dust holes, others may not; we therefore decided to use a stricter definition of the transition disk so as not to include anaemic IRAC disk sources. We return to this discussion in section \S\, \ref{subsec:evolution}.

Well-known examples of sources with transition disks include TW\,Hya (10\,Myr) \citep{calvet02,calvet05}, GM\,Aur (2-3Myr) \citep{calvet05, grafe11}, DH\,Tau, and DM\,Tau \citep[e.g.][]{grafe11}. Transition disks have holes that are typically a few AU ($\leq$ 10) in radius, and display a broad H$\alpha$ emission line. The spectroscopic study by \citet{furesz08} of sources in the Orion Nebula Cluster revealed sources that have no detectable disk emission short-wards of 8\um\ yet show clear H$\alpha$ accretion signatures. These authors found approximately 35 sources with these characteristics, interpreting them as sources with transition disks. The age at which these sources may be found can therefore be as young as $\sim$1\,Myr. 

Sources with no IRAC excess and H$\alpha$ equivalent widths $\ge$\,10\,\AA\ might also be dMe stars masquerading as sources with transition disks. 
Main-sequence M-type stars are classified as either having high (dMe) or low (dM) chromospheric and coronal activity; the magnetic activity is identified via H$\alpha$ emission, i.e., dMe stars have typically equivalent widths 15$\ge$\,EW(H$\alpha$)\,$\ge$1\,\AA , while dM stars have EW(H$\alpha$)\,$<$1\,\AA\ \citep[e.g.][]{hawley96,kruse10}. In addition, dMe stars have Ca\,II HK resonance doublet lines in emission, as well as X-ray flaring. dMe stars may therefore be mistaken for either pre-main sequence WTTS or transition disks. 
\citet{riaz06b} analysed \emph{Spitzer} IRAC and MIPS data for dMe stars and found that their SEDs could be described by a purely photospheric spectrum out to 24\um, so longer wavelength data, i.e. $\ge$\,24\um , is necessary to identify bona-fide transition disk sources. 
\citet{dahm05a} discussed the contamination of dMe stars in their H$\alpha$ survey of NGC2264 and estimated it to be $\sim$8\%; from their analysis, it is expected that foreground dMe stars are very faint and red in the (V-I) vs. I optical colour-magnitude diagram; after examining optical photometry, we found that our sources do not populate this part of the diagram and are thus unlikely to be dMe stars.


\section{Discussion}
\label{sec:dis}

\subsection{Star formation efficiency}

As previously mentioned, the FLAMINGOS NIR dust extinction map indicates that the molecular cloud associated with the cluster has mean and median A$_\mathrm{V}$ of 5.9 and 4.3\,mag, respectively. We built a 2MASS control field $K_s$-band luminosity function (KLF) and reddened it by these two values, creating two reddened KLFs, and subsequently subtracted each one of them from the FLAMINGOS KLF to statistically estimate the number of sources in excess that would thus be members of the NGC2264 cluster. We opted for calculating the population using both the mean and median values stated above, for these correspond to the lower and upper limits of the real value.

\begin{figure}[!h]
\centering
\includegraphics[width=0.48\textwidth]{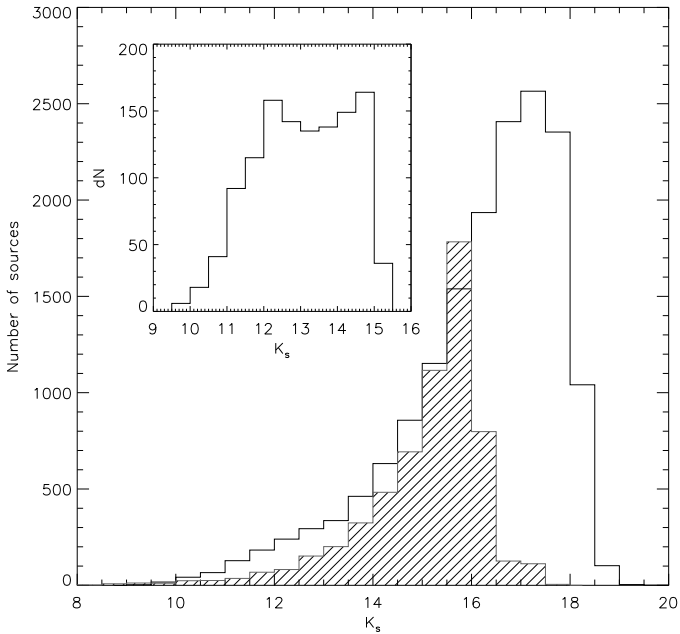}
\caption{Comparison of the NGC\,2264 FLAMINGOS $K_s$-band luminosity function (KLF) (solid black line) to the reddened 2MASS control field KLF (red hatched-line histogram). The histogram in the inset shows the differential KLF, with 1194 sources.}
\label{fig:dklf}
\end{figure}

Figure \ref{fig:dklf} shows two KLFs ( the FLAMINGOS KLF and the 2MASS KLF uniformly reddened by A$_\mathrm{V}$=4.3\,mag), as well as the corresponding differential KLF. The number of sources of the differential KLF is 1194, for $K_s< $15.5\,mag (corresponding to a 0.2\,M$_\odot$ star, using \citet{baraffe98} PMS evolutional models, for a distance of 913\,pc and an age of 2.5\,Myr). Using the same procedure, but uniformly reddening the 2MASS KLF instead by A$_V$=5.9\,mag yields a population of 1677 sources ($K_s < $15.5\,mag) within the same area of 1264.48 arcmin$^2$. The size of the stellar population of the NGC2264 cluster (within this area) is therefore estimated to be 1436$\pm$242. \citet{lada93} used this same method to statistically determine the number of sources of NGC\,2264: they found 360$\pm$130 sources ($K_s<$ 13.5\,mag), albeit for a smaller area: 540 arcmin$^2$.

Assuming, for simplicity, that each source has 0.5\,M$_\odot$  \citep{muench07}, then the stellar mass of NGC2264 is 718$\pm$121\,M$_\odot$. This value can be used to estimate the star formation efficiency (SFE) of NGC\,2264, which is defined as SFE=stellar mass/(stellar mass + cloud mass). For the cloud mass, we used the mass measured from regions with extinction greater than the star formation extinction threshold, i.e.,  A$_V \ge$7\,mag \citep{johnstone04, lada10}, (2.1$\pm 0.5) \times$10$^3$\,M$_\odot$. The star formation efficiency of the cloud is calculated to be approximately 25\,\%, similar to the value determined by \citet{lada03} for nearby star forming clusters. A more accurate estimate of the total stellar mass can of course be obtained by using an appropriate IMF, although that is beyond the scope of this paper. The reader should also note that the statistically determined members are all brighter than $K_s$=15.5\,mag, i.e. the fainter (deeply embedded young protostellar and/or lower mass) members are not accounted for and so the stellar mass used is an underestimate.

\subsection{Disk evolution}
\label{subsec:evolution}

A new empirical evolutionary paradigm is emerging for primordial optically thick disks, which consists of two distinct pathways \citep[e.g.][and references therein]{lada06a,najita07b, cieza07b, cieza08,merin08,currie09, cieza10}. In one of these evolutionary paths that a primordial disk can take, it experiences a homologous depletion of dust \citep[via dust settling and grain growth,][]{wood02a} throughout the entire extent of the disk; this path is characterized observationally by SEDs with slopes that do not vary substantially longwards of 3.6\um. The alternative evolutionary path a primordial disk can undergo consists of a radial, inside-out clearing of the dust disk. The SEDs of sources that follow this latter path have different slopes shortwards and longwards of 8\um. The results presented in this paper lend further support for this empirical disk evolution paradigm.

We showed in section \S\,\ref{subsubsec:disk-av} that the anaemic disk population differs from that of the thick disks in two main aspects: its spatial distribution displays no segregation, and it is an unembedded population (A$_V <$3\,mag). Both of these facts indicate that the anaemic disk population is more evolved than the thick disk population, having had enough time to emerge from the parental molecular cloud and begin populating the halo of the NGC\,2264 cluster.

\begin{figure}[!h]
\centering
\includegraphics[width=0.48\textwidth]{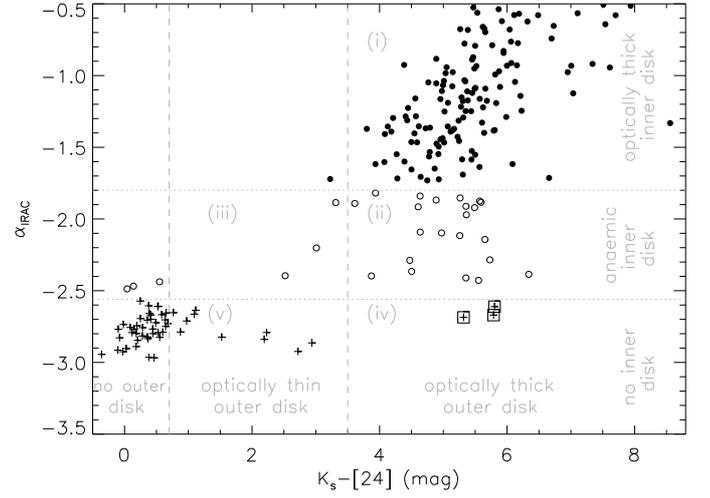}
\caption{($K_s$-[24]) vs. \airac\ diagram for sources with no IRAC disk (plus symbols), sources with IRAC anaemic inner disks (open circles), and sources with thick inner IRAC disks (filled circles). The plot is divided into nine sectors by grey dotted horizontal lines and vertical dashed lines \citep[following][]{cieza08}; see the discussion in \S\,\ref{subsec:evolution} for a detailed description of sectors (i) to (v). Sources that were identified as candidate transition disks from Figure \ref{fig:airac-halpha} are additionally marked by open squares.}
\label{fig:airac-mips}
\end{figure}

Figure \ref{fig:airac-mips} characterizes the outer disk for the two aforementioned populations, using MIPS 24\um\ data. The ($K_s$-[24]) colour is used as a proxy for estimating the optical thickness of the outer disk  \citep[following Figure 2 of][]{cieza08}. We did not detect all of the IRAC sources at 24\um: only 29\% of the anaemic sources have 24\um\ counterparts, while 63\% of the thick disk sources were detected at 24\um. The 24\um\ sources that are most likely detected are those with large excesses due to the cluster's distance, and indeed, Figure \ref{fig:airac-mips} shows that the IRAC disk sources detected at 24\um\ practically all have optically thick outer disks. The diagram identifies sources with (i) optically thick inner and outer disks,  (ii) anaemic inner disk and thick outer disk, (iii) anaemic inner disk and optically thin outer disk, (iv) sources with no inner disk and optically thick outer disks, and (v) sources with no inner disk and optically thin outer disks. 

The figure shows the path for a radial or inside-out  evolution of primordial optically thick disk, since we have identified sources in all the different phases.  The younger sources could evolve from (i) to (ii) and finally to transitional disks (iv). On the other hand, a homologous disk evolution could be described by young sources in (i), evolving into (iii). As explained before, region (iii) of the diagram is not well-populated by sources because the 24\um\ detections are biased towards large excesses. Taking into account that ~70\% of the anaemic inner disks do not have 24\um\ counterparts, we can reasonably assume that their outer disks are optically thin. These results may therefore indicate that the majority of primordial optically thick disks tend to follow  homologous depletion evolutionary path. 

Multiple evolutionary paths have also been observationally identified in other young star-forming regions, such as IC\,348 \citep{lada06a}, Lupus, Serpens, and Cha\,II \citep{merin08}. These regions are all much closer than NGC\,2264, thus do not suffer from the detection bias at 24\um. \citep{lada06a} compiled an extensive atlas of SEDs for disk sources and found that the most numerous type of disk was one that can be described by homologous depletion. Furthermore, studies of NGC\,2362 \citep{currie09} and the Coronet cluster \citep{currie11} also show that the frequency of homologously depleted disks is higher than that of disks with inner holes.
On the other hand, \citet{cieza10} carried out a detailed study of evolved disks in the nearby Ophiucus molecular cloud, and argue that at one point during their evolution all disk sources pass  through a phase where the outer disk remains optically thicker than the inner disk. 

\subsubsection{Transition disks}
 There are currently several possible scenarios that could account for the formation of holes within protoplanetary disks. The first of these clearing processes is (a) photoevaporation and viscous disk evolution processes \citep[e.g.][]{alexander07a}. \citet{alexander06a,alexander06b} conducted a theoretical study of disk evolution, using a model combining both viscous evolution of the disk and photoevaporation of the disk by stellar radiation, and suggested that all evolving disks pass through a short inner hole phase; they furthermore predict that sources with transition disks should represent approximately 10\% of the disk sources. Another process that can clear an inner dust hole is (b)  grain growth and planet formation \citep[e.g.][]{tanaka05,flaherty08}. \citet{dullemond05} predict, using evolutionary models, that grain growth and dust settling occurs more rapidly in the inner region of the disk than in the outer regions. Inner disk holes can also be formed through the (c) dynamical interaction of the disk with an embedded giant planet \citep[e.g.][]{alexander09}. Finally, the process of (d) dynamical interaction of the disk with a (sub)stellar companion \citep[e.g.][]{ireland08} may also remove dust from the inner part of the disk. 

We identified four candidate transition disks from Figure \ref{fig:airac-halpha}; these sources are accreting according to their H$\alpha$ equivalent width and do not possess IRAC inner disks. We indicated three of these sources in Figure \ref{fig:airac-mips} with squares (the fourth candidate has no 24\um\ detection), and the diagram shows they all have optically thick outer disks.

 If we consider these four candidate transition disks, then the fraction of transition disks for NGC\,2264 is $\sim$4\% (for a total 105 disked sources with H$\alpha$ measurements).  Considering the 24\um\ detection bias in our data, our transition disk fraction may well be an underestimate. This value is lower than that found in other studies,  such as \citet[e.g.][]{sicilia-aguilar06} who found that 10\% of the disked sources in the young cluster Trumpler 37 are sources with transition disks. This value is in accord with \citet{alexander06a,alexander06b}'s proposal that 10\% of the disked sources are transition disks. A similar value was proposed by \citet{butler06}, namely that $\sim$12\% of FGKM stars host giant planets within 30\,AU, implying that giant planet formation clears the inner disk.


\section{Summary}

We have found that sources with anaemic inner disks are more evolved than those with thick inner disks. We presented several results that lead to this conclusion, namely:
   \begin{enumerate}
      \item Sources with thick disks are mostly embedded, while sources with anaemic disks are mostly unembedded. Using the NIR+IRAC colour-colour diagrams and their disk loci, we determined the extinction towards sources with anaemic and thick disks: $\sim$20\% of the sources with thick disks are embedded (A$_\mathrm{V} >$  3\,mag), while  none of the sources with anaemic disks are embedded (A$_\mathrm{V} >$  3\,mag).
      \item The spatial distribution of extincted sources with thick disks or sources with anaemic disks is very different: extincted sources with thick disks are found to group into three clusterings (near IRS\,1, the Spokes cluster, and in the Fox Fur Nebula), whereas sources with anaemic disks are more spread-out throughout the region with no clear clustering, populating the larger halo of the cluster. This indicates that sources with anaemic disks have had time to move away from their birth-site and expand into the whole region. 
      \item Most of the sources with thick disks have H$\alpha$ emission that is consistent with accretion namely $\sim$85\%, while only $\sim$59\% of the sources with anaemic disks appear to be accreting.
   \end{enumerate}
Our MIPS 24\um\ observations of the cluster detected 29\% of the anaemic disk sources, and 63\% of the thick disk sources. Unfortunately, our MIPS observations are very likely biased toward identifying sources with large excesses at 24\um\ due to the distance of the cluster. Most of the detected disk sources have optically thick outer disks.
In addition, we identified four candidate transition disk sources in NGC\,2264. These sources are unembedded, appear to be accreting, and in three cases have optically thick outer disks.  We also found that the sources with candidate transition disks are more evolved than sources with thick disks. 

Our findings support the emerging disk evolution paradigm of two distinct evolutionary paths: one where the disk is homologously depleted throughout its radial extent, and another where the disk evolves radially, from the inside-out. Since there are many more anaemic disks in NGC\,2264 than transition disks, the homologous depletion disk evolutionary path appears to be the one that most primordial optically thick disks follow. 

Regarding the star formation history of the cluster, the spatial distribution of sources indicates that NGC\,2264 may have undergone more than one star-forming event. The star formation may have initiated in the north, near S\,Mon, and continued to progress toward the south where it is forming stars now in the Spokes cluster and in the vicinity of Allen's source. We also identified a potential triggering of star formation in the north-west of the cluster, in the Fox Fur Nebula.

Finally, we calculated the mass of the dense molecular cloud associated with the cluster to be 2.1$\times$10$^3$\,M$_\odot$ (for A$_\mathrm{V} >$ 7\,mag). We also estimated the total population of the cluster to be 1436$\pm$242 stars, and a star formation efficiency of $\sim$25\%.

\begin{acknowledgements}

We warmly thank Luisa Rebull for sharing her ancillary data catalogue of NGC\,2264 with us. We also thank Naomi Ridge for kindly allowing us to directly compare her $^{13}$CO and C$^{18}$O observations of NGC\,2264 with our data.
P.~S.~T. acknowledges support from the grant SFRH/BD/13984/2003 awarded by the Funda\c{c}\~ao para a Ci\^encia e Tecnologia (Portugal).
This work is based on observations made with the {\emph Spitzer} Space Telescope, which is operated by the Jet Propulsion Laboratory, California Institute of Technology under NASA contracts 1407 and 960785. 
FLAMINGOS was designed and constructed by the IR instrumentation group (PI: R.~Elston) at the University of Florida, Department of Astronomy, with support from NSF grant AST97-31180 and Kitt Peak National Observatory. The FLAMINGOS data were collected under the NOAO Survey Program, \emph{Towards a Complete Near-Infrared Spectroscopic Survey of Giant 
Molecular Clouds} (PI: E. Lada) and  supported by NSF grants AST97- 3367 and AST02-02976 to the University of Florida. this publication is based in part upon work supported by the National Science Foundation under Grant No. AST-1109679 and NASA grant LTSA NNG05D66G awarded to the University of Florida.
This publication makes use of data products from the Two Micron All Sky Survey, which is a joint project of the University of Massachusetts and the Infrared Processing and Analysis Center/California Institute of Technology, funded by the National Aeronautics and Space Administration and the National Science Foundation.

\end{acknowledgements}


\clearpage

\clearpage

\begin{table*}[!h]
\caption{Summary of the FLAMINGOS observations.\label{tab:flamingos}}
\centering
\begin{tabular}{cccccccl}
\hline\hline
Field & R.A. (J2000) & Dec. (J2000) & Passband & Dithers & Exp. time(s) & Total exp. time (s) & Observation date\\
\hline
A-3  & 06\hr41\m31.33\s & +09\degr30\arcmin08.2\arcsec & $J$  & 48 & 60.0 & 2304 & 27 Dec 2002, 31 Jan 2003, 01 Feb 2003\\ 
     &                  &                              & $H$  & 14 & 60.0 & 840  & 31 Jan 2003\\
     &                  &                              & $K_s$ & 31 & 35.0 & 1085 & 31 Jan 2003\\
\hline
A-4  & 06\hr41\m31.33\s & +09\degr50\arcmin08.2\arcsec & $J$  & 32 & 60.0 & 1920 & 27 Dec 2002, 01 Feb 2003\\
     &                  &                              & $H$  & 32 & 60.0 &      & 27 Dec 2002, 01 Feb 2003\\
     &                  &                              &     & 32 & 30.0 & 2880 & 27 Dec 2002\\
     &                  &                              & $K_s$ & 30 & 35.0 & 1050 & 01 Feb 2003\\
\hline
A-12 & 06\hr40\m10.22\s & +09\degr30\arcmin08.2\arcsec & $J$  & 14 & 60.0 & 840  & 03 Jan 2003\\
     &                  &                              & $H$  & 15 & 60.0 & 900  & 03 Jan 2003\\
     &                  &                              & $K_s$ & 14 & 60.0 & 840  & 22 Mar 2003\\
     &                  &                              &     & 32 & 35.0 & 1120 & 23 Mar 2003\\
\hline
A-13 & 06\hr40\m10.22\s & +09\degr50\arcmin08.2\arcsec & $J$  & 16 & 60.0 & 960  & 23 Mar 2003\\
     &                  &                              & $H$  & 16 & 60.0 & 960  & 03 Jan 2003\\
     &                  &                              & $K_s$ & 16 & 60.0 & 960  & 22 Mar 2003\\
     &                  &                              &     & 36 & 35.0 & 1260 & 23 Mar 2003\\
\hline
c1   & 06\hr40\m53.21\s & +09\degr51\arcmin08.0\arcsec & $J$  & 32 & 60.0 & 1920 & 30 Jan 2003, 30 Jan 2004\\
     &                  &                              & $H$  & 32 & 60.0 & 1920 & 28 Jan 2003, 30 Jan 2004\\
     &                  &                              & $K_s$ & 66 & 30.0 &      & 28 Jan 2003\\
     &                  &                              &     & 32 & 35.0 & 3100 & 29 Jan 2003, 30 Jan 2004\\
\hline
c2   & 06\hr41\m10.82\s & +09\degr31\arcmin06.8\arcsec & $J$  & 47 & 60.0 & 2820 & 28 Jan 2003, 01 Jan 2005\\
     &                  &                              & $H$  & 16 & 60.0 & 960  & 28 Jan 2003\\
     &                  &                              & $K_s$ & 64 &
     30.0 & 1920 & 28 Jan 2003, 01 Jan 2005\\
\hline\hline
\end{tabular}
\end{table*}

\begin{table}[!h]
\caption{Source classification scheme using \airac.\label{tab:class}}
\centering
\begin{tabular}{lc}
\hline\hline
Source classification      & \airac\ value \\
\hline
Class\,I (C\,I) sources               & \airac $\ge$ 0.5\\
Flat spectrum (FS) sources            & 0.5 $>$ \airac $\ge$ -0.5\\
Sources with thick disks (TD)         & -0.5 $>$ \airac $\ge$ -1.8\\
Sources with anaemic disks (AD)        & -1.8 $>$ \airac $\ge$ -2.56\\
Sources with naked photospheres (NP)  & \airac $\le$ -2.56 \\
\hline
\end{tabular}
\end{table}

\begin{table}[!h]
\caption{Coefficients for Equation \ref{eq:loci}}
\centering
\label{tab:loci}
\begin{tabular}{ccc}
\hline\hline
Colour-colour space & a$_0$ & a$_1$ \\
\hline
($K_s$-[8]) vs. ($J-H$)   &  0.56$\pm$0.04 &  0.10$\pm$0.02\\
($K_s$-[5.8]) vs. ($J-H$)  & 0.47$\pm$0.03 &  0.25$\pm$0.03\\
($K_s$-[4.5]) vs. ($J-H$)   & 0.634$\pm$ 0.12 & 0.170$\pm$ 0.09\\
($K_s$-[3.6]) vs. ($J-H$)   &  0.40$\pm$0.13 &  0.65$\pm$0.10\\
\hline
\end{tabular}
\end{table}

\clearpage 
\bibliographystyle{aa}

\clearpage
\appendix
\section{Modelling the disk locus}

\begin{figure}[!h]
\centering
\includegraphics[width=0.48\textwidth]{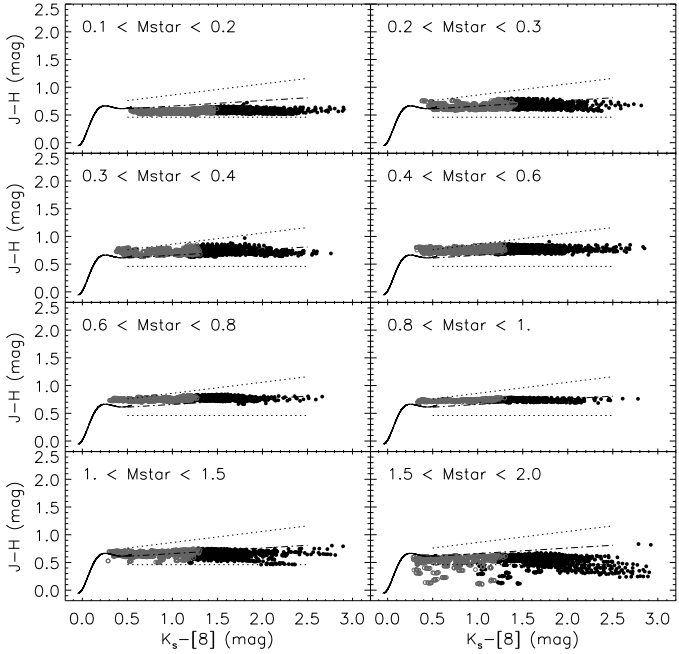}
\caption{Color-color diagrams using models from \citet{robitaille06} for disked sources with inclinations smaller than 75\degr\ for different ranges of stellar masses (M$_\odot$). The black solid line represents the Pleiades main sequence, the dot-dashed line corresponds to the disk locus, and the dotted lines represent the 5-$\sigma$ error loci as determined by the observations of NGC\,2264 (see \S\,\ref{subsubsec:disk-av}, Figure \ref{fig:hess} and Table \ref{tab:loci}). The black dots represent sources with optically thick disks, and the grey circles represent sources with anaemic disks.}
\label{fig:appendixf1}
\end{figure}

\begin{figure}[!h]
\centering
\includegraphics[width=0.48\textwidth]{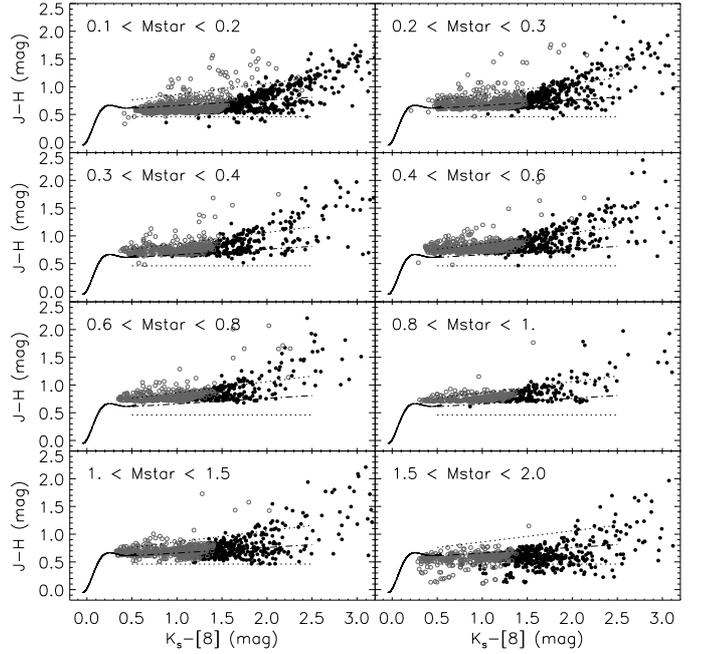}
\caption{Same as Fig. \ref{fig:appendixf1}, except for models of disks with inclination angles greater than 75\degr. }
\label{fig:appendixf2}
\end{figure}

We discussed in section \S\,\ref{subsubsec:disk-av} empirical disk loci for NGC\,2264. To better understand how the disk locus depends on stellar and disk parameters, we used radiation transfer  models of disk sources from \citet{robitaille06} to explore the ($K_s-[8]$) vs. ($J-H$) colour-colour diagram in more detail. Figure \ref{fig:appendixf1} uses models with disk inclinations $\leq$70\degr\ and varying stellar mass. The modelled sources are classified according to the SED slope into sources with either optically thick disks (black symbols) or anaemic disks (grey symbols), in the same manner as we did for our data (Table \ref{tab:class}).  For comparison, we overplot the empirically determined disk locus and 5-$\sigma$ limits from Table \ref{tab:loci} on each diagram. We found no significant variation in the locus for sources with disks with varying inclinations angles below 75\degr, although the locus does have significant dependence on the stellar mass in this colour-colour space. Figure \ref{fig:appendixf2} shows the same set of modelled sources as Fig. \ref{fig:appendixf1}, but with inclination angles $>$ 75\degr\ (i.e., closer to edge-on disks). These plots show us that the locus is not as well-constrained, and the sources fan out from the locus; this effect is stronger for sources with thick disks. 

\begin{figure}[!h]
\centering
\includegraphics[width=0.48\textwidth]{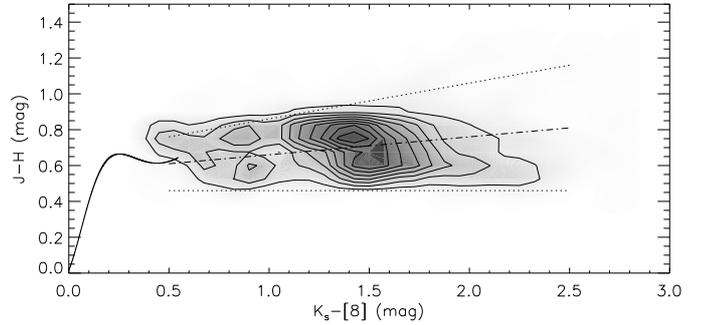}
\caption{Synthetic color-color diagram built using the models shown in Fig. \ref{fig:appendixf1} and Fig. \ref{fig:appendixf2} with an IMF sampling \citep{kroupa01,robitaille06}. The solid line represents the main sequence, the dot-dashed line the empirical disk locus, and the dotted lines the 5-$\sigma$ error loci discussed in \S\,\ref{subsubsec:disk-av}, and shown in Figure \ref{fig:hess} and Table \ref{tab:loci}.}
\label{fig:appendixf3}
\end{figure}

To try and reproduce the empiral disk locus, we built a synthetic cluster using the probabilities calculated by \citet[][i.e., assuming a Kroupa initial mass function between 0.1 and 30\,M$_\odot$, and with stellar ages sampled linearly in time between 1000\,yr and 2\,Myr, and random disk inclination angles]{robitaille06}. Figure \ref{fig:appendixf3} shows the resulting ($K_s-[8]$) vs. ($J-H$)  colour-colour diagram. This plot shows that it is impossible to \emph{accurately} determine extinctions of sources using solely the disk locus because it has a typical intrinsic width in ($J$-$H$) colour of $\pm$0.3\,mag (for a strong 5-$\sigma$ cut), which translates to an A$_V$ of 2.7\,mag. However, the diagram is useful for robustly identifying embedded sources and unembedded sources, as we demonstrated in section \S\,\ref{subsubsec:disk-av}.

\end{document}